\documentclass[twocolumn,epsf,prb,amsmath,amssymb,showpacs]{revtex4}
\usepackage{graphicx}

\begin{document}
\title{Magnetic interface states in graphene-based quantum wires}
\author{J. Milton Pereira$^{1,2}$ Jr., F. M. Peeters$^1$,P. Vasilopoulos$^3$}
\address{$^1$Department of Physics, University of Antwerp, Groenenborgerlaan 171, B-2020 Antwerpen\\
$^2$Departamento de F\'{\i}sica, Universidade
Federal do Cear\'a, Fortaleza, Cear\'a, $60455$-$760$, Brazil\\
$^3$Department of Physics, Concordia University, Montreal, Quebec,
Canada H3G 1M8}

\begin{abstract}
The electronic states of a finite-width graphene sheet in the
presence of an electrostatic confining potential and a
perpendicular magnetic field are investigated. The confining
potential shifts the Landau levels inside the well and creates
current-carrying states at or close to the interface with the
barriers in addition to the edge states caused by the finite width
of the sheet. Detailed energy spectra are given as a function of
the quantum wire parameters. The dependence of the density of
states on the confinement potential is evaluated for finite and
zero magnetic field.
\end{abstract}
 \maketitle


\section{Introduction}
The recent production of single layers of stable carbon crystals
\cite{Novo3,Novo2,Zhang1} has attracted a large interest in their
fundamental properties and their potential technological
applications. The unusual properties of carriers in graphene are a
consequence of the gapless and approximately linear electron
dispersion at the vicinity of the Fermi level at two inequivalent
points of the Brillouin zone. In the low-energy limit the
quasiparticles in these systems are described in terms of massless
chiral relativistic fermions governed by the Dirac equation. In
particular, graphene has been shown to display an unusual quantum
Hall effect \cite{Zheng,Novoselov,Sharapov,Zhang}, in which the
quantum Hall plateaus are found in half-integer multiples of 4.
This results from the fourfold degeneracy of the Landau levels
(LL) along with the existence of a non-zero Berry phase. The edge
states in graphene in a perpendicular magnetic field were also
found to display unusual properties, such as counterpropagating
spin-polarized modes \cite{Abanov}, and are expected to play a
particularly important role in thin graphene structures.

Another important result concerning single graphene layers is the
possibility of controlling the electron density and Fermi level by
a gate voltage. This allows the creation of graphene-based quantum
structures such as potential barriers and quantum wires (QW).
Theoretical studies have shown that the relativistic behavior of
quasiparticles in graphene allow the observation of effects such
as the Klein paradox, which is the perfect transmission of
relativistic particles across potential barriers, as well as a
direction-dependent tunneling through barriers
\cite{Klein,Milton,kat,Falko,Milt2}. In addition, recent
experimental work has demonstrated electronic confinement in
patterned graphene structures created by standard lithography
methods \cite{Berger}.

In this paper we study the interplay of an electrostatic potential
barrier and an external perpendicular magnetic field on a graphene
QW and find that propagating states exist at the interface with
the potential barriers in addition to the ordinary edge states.
Further, for sufficiently wide potential wells two distinct sets
of LL arise due to the energy shift caused by the presence of
barriers.

This paper is organized as follows. In Sec. II we present the
model and formalism and in Sec. III numerical results.  We
conclude with remarks in Sec. IV.

\section{Model and formalism}
The crystal structure of undoped, defect-free graphene layers is
that of a honeycomb lattice of covalent-bond carbon atoms. To each
carbon atom corresponds a valence electron and the structure can
be described as composed of two sublattices, labelled A and B. The
low-energy excitations of the system at the vicinity of the
${\mathbf K}$ point and in the presence of both an electrostatic
potential $U$ and a uniform magnetic field $B$ perpendicular to
the plane of the graphene sheet are described, in the continuum
approximation, by the 2D Dirac equation
\begin{equation}
\{v_F[\vec{\sigma}\cdot \hat{\mathbf p}-e{\mathbf A}]+m\, v_F^2
\sigma_z\} \Psi = (E-U)\Psi,
\end{equation}
where the pseudospin matrix $\vec {\sigma}$ has components given
by Pauli's matrices; $\hat{\mathbf p} = (p_x,p_y)$ is the momentum
operator.  The "speed of light" of the system is $v_F$,
 the Fermi velocity ($v_F \approx 1\times 10^6$ m/s), and
${\mathbf A}$ is the vector potential. The eigenstates of Eq. (1)
are represented by two-component spinors $\Psi = [\psi_A \, , \,
\psi_B]^T$, where $\psi_A$ and $\psi_B$ are the envelope functions
associated with the probability amplitudes at the respective
sublattice sites of the graphene sheet. The term $\propto m\,
v_F^2$ introduces an energy gap, which may represent e.g. the
effect of spin-orbit coupling.

We now consider a narrow graphene layer, of width $W$, in the
presence of a one-dimensional (1D) potential $U=U(x)$ and a
perpendicular magnetic field $B$. This allows us to write the
solutions for the spinor components in the form
$\psi_A(x,y)=\phi_A(x)e^{ik_y y}$ and
$\psi_B(x,y)=i\phi_B(x)e^{ik_y y}$ because of translational
invariance along the $y$ direction and the particular choice of
Landau's gauge ${\mathbf A}=(0,Bx,0)$. The resulting equations for
$\phi_A(x)$ and $\phi_B(x)$ are
\begin{eqnarray}
\frac{d\phi_A}{d x} - (k_y-eB x)\phi_A &=&
-[E-U(x)+m\,v_f^2]\phi_B,\cr &{}&\cr \frac{d\phi_B}{d x} + (k_y-eB
x)\phi_B&=&[E-U(x)-m\,v_f^2]\phi_A.
\end{eqnarray}
These equations can be decoupled and, by setting $\xi =
\beta^{1/2}(x - k_y /\beta)$,  where $\beta = \ell_B^{-2} =
eB/\hbar$ is the inverse magnetic length squared, the result is
\begin{eqnarray}
d^2 \phi_A/d \xi^2 &+& [(\Omega + 1) - \xi^2]\phi_A \cr &-&
\frac{u'}{(\epsilon - u + \Delta)}[d \phi_A/d \xi + \xi\phi_A]= 0,
\end{eqnarray}
\begin{eqnarray}
d^2 \phi_B/d \xi^2 &+& [(\Omega - 1) - \xi^2]\phi_B \cr &+&
\frac{u'}{(\epsilon - u - \Delta)}[d \phi_B/d \xi - \xi\phi_B]= 0,
\end{eqnarray}
where $\Omega = [(\epsilon - u)^2-\Delta^2]/\beta$, $u = (\hbar
v_F)^{-1}U$, $\epsilon = (\hbar v_F)^{-1}E$, $\Delta =
m\,v_F/\hbar$ and the prime denotes derivative with respect to
$\xi$. For a constant potential $U=U_0$, Eqs. (2) and (3) have
well-known solutions in terms of Hermite polynomials and the
spectrum is
\begin{equation}
E = \pm\hbar v_F\sqrt{2n\beta + \Delta^2}+U_0,
\end{equation}
where $n$ is an integer. This contrasts significantly with the
nonrelativistic spectrum  $E = \hbar\omega_c(n+1/2)+U_0$.

\section{Results}
\subsection{Dispersion relation and density of states}

First we consider the effect of a steep potential well $U(x)$,
with a characteristic width $L$ in a graphene strip of width $W$.
In this case the derivatives of the potential are strongly
localized functions that have non-zero values only at the vicinity
of the barrier interfaces. The solutions then depend on the width
$W$ and the strength of the magnetic field $B$ through the ratio
$\ell_B/L$. Let us consider initially the case $2\ell_B/L << 0.5$,
which corresponds to classical orbits that fit inside the QW. We
can assume solutions of the form $\phi_C = f_C(\xi)e^{-\xi^2/2}$,
$C=A,B$. For a constant potential $f_C$ are the well-known Hermite
polynomials. Thus, for sufficiently strong $B$ and small $k_y$ the
spinor functions quickly decay with $\xi$ and the solutions are,
to a good approximation, localized inside the well. Therefore, the
energy spectrum is dispersionless and for small $n$ is that of the
LL for $U=0$. For larger values of $k_y$ the center of the
solutions $\ell_B^2k_y$ is shifted towards the barrier regions.
For the lowest LL ($n = 0$) one can estimate the limits of the
central dispersionless region by setting
$\exp(-\xi^2/2)\approx
0.1$, or $\xi \approx 2$,
since in this case the amplitude of the
wave function inside the barriers is negligible. This gives $-1+
4\ell_B /L < 2k_y \ell_B^2/L <1-4\ell_B /L$. For non-zero values
of $n$, the central dispersionless region of the spectrum is
expected to be narrower, since the spinor functions are less
localized and can have a larger magnitude within the barrier
regions.

For sufficiently large values of the momentum along the $y$
direction, a similar argument shows that the energy levels may be
accurately approximated by the LL shifted by $U_0$ in the regions
$2k_y \ell_B^2/L <-1-4\ell_B /L$ and $2k_y \ell_B^2/L
>1+4\ell_B /L$. For intermediate values of $k_y$ one expects
dispersive solutions. These solutions can be described as
interface states, in the sense that they are localized electronic
states in the $x$ direction, that propagate along the interfaces
with the potential barriers, in analogy with the edge states of a
2D electron gas in a magnetic field, but with the fundamental
difference that in the present case the spinor functions are
non-negligible both inside the QW and in the barriers. These are
current-carrying states and, for sufficiently smooth potentials,
they should not depend on the microscopic structure of the
graphene sheet. For even larger values of $k_y$ the spinor
functions may be shifted toward the edges of the sample and give
rise to edge states, which have been shown to depend on the shape
of the graphene edges \cite{Brey}.

We have considered two specific types of potential wells, as shown
in Fig. 1. The solid line illustrates the step potential given by
\begin{equation}
U(x) =
\frac{U_0}{2}\{\tanh[(-x-\frac{L}{2})/\delta]+\tanh[(x-\frac{L}{2})/\delta]+2\},
\end{equation}
where $\delta$ denotes the thickness of the interface. The dashed
line in Fig. 1 refers to a parabolic potential
\begin{equation}
U(x)=\left\{
\begin{array}{ccc}
U_0(2x/L)^2 && |x| < L/2, \\
&&\\
U_0 && |x| > L/2.
\end{array}
\right.
\end{equation}
Figure 2 shows numerical results for the energy spectrum of a QW
with the potential given by Eq. (6). The results are for $L =200$
nm and an interface width of $10$ nm, which is much larger than
the lattice parameter of graphene $a \approx 0.14$ nm, $B = 2$ T
and $U_0 = 50$ meV. In this case we have $2\ell_B/L \approx 0.18$.
The vertical dashed lines delimit the range $1-4\ell_B /L<|2k_y
\ell_B^2/L| <1+4\ell_B /L$. In contrast with the conventional edge
states, the barrier interface does not cause a splitting of the
dispersion branches. This is a consequence of the difference in
boundary conditions at the potential interfaces and the edges of
the graphene sheet, i.e., at the potential step the wave function
is finite and continuous and the associated probability density
can be significant for $|x| > L/2$, whereas at the edges of the
sample, as in the case of a graphene sheet with an armchair
termination, the wave function was assumed to vanish.

\begin{figure}
\centering{\resizebox*{!}{7.5cm}{\includegraphics{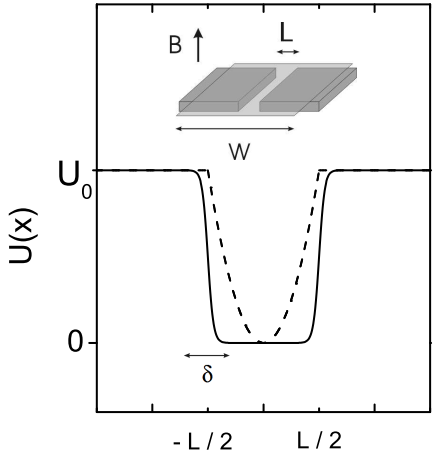}}}
\vspace{-0.5cm} \caption{Schematic depiction of the different
potential well profiles discussed in the text: tangent hyperbolic
(solid line), parabolic  (dashed line).} \label{fig:f1}
\end{figure}
\begin{figure}
\centering{\resizebox*{!}{9.0cm}{\includegraphics{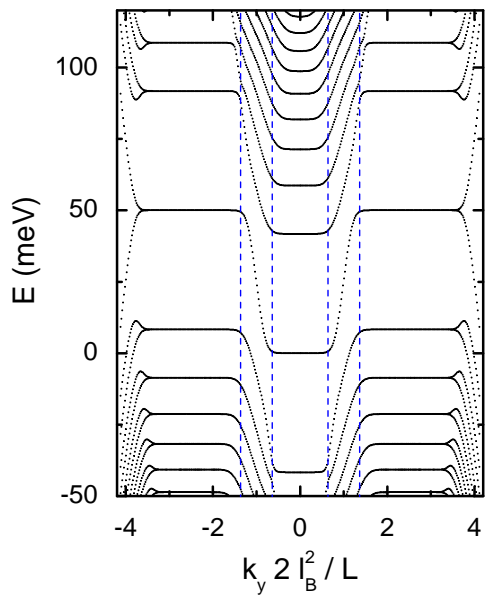}}}
\vspace{-0.5cm} \caption{Energy spectrum of a graphene QW for $B =
2$ T and a potential given by Eq. (6), with $U_0 = 50$ meV, $L =
200$ nm} \label{fig:f2}
\end{figure}
\begin{figure}
\centering{\resizebox*{!}{9.0cm}{\includegraphics{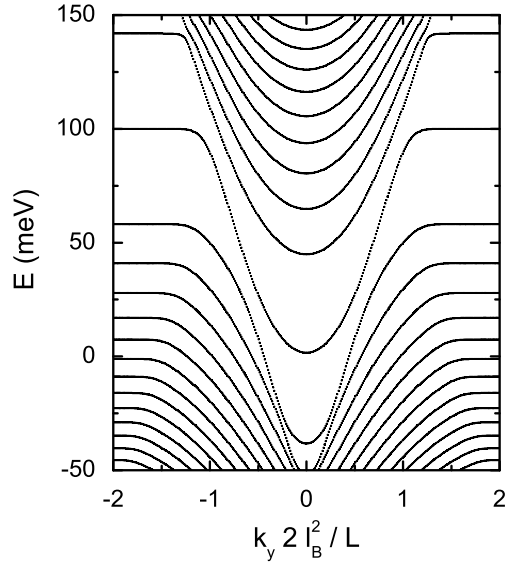}}}
\vspace{-0.5cm} \caption{The same as Fig. 2 but for a parabolic
potential  given by Eq. (7), with $U_0 = 100$ meV, $L = 200$ nm,
$B = 2$ T.} \label{fig:f3}
\end{figure}
For comparison, Fig. 3 shows the spectrum of a parabolic confining
potential given by Eq. (7). This potential is non-zero at every
point, except at $x = 0$, and is the reason why the states are
dispersive for all values of $k_y$ satisfying $|2k_y \ell_B^2/L|<
1$. This result becomes similar to that of the tangent hyperbolic
potential for large $\delta$.

For weaker fields, such that $2\ell_B/L > 0.5$, the former picture
breaks down and the magnetic field acts as a perturbation to the
zero-field case \cite{Milton}. For $B = 0$ the electron states
inside the QW that propagate perpendicularly to the barrier
interfaces are transmitted without reflection (Klein
tunnelling)\cite{Klein,kat,Falko,Milton}. This counterintuitive
behavior results from the absence of a gap in the spectrum and
from the chiral nature of the quasiparticles in graphene. However,
recently it has been demonstrated  that non-zero values of
momentum along $y$ allow the existence of confined electron states
in a QW \cite{Milton}. For large values of $k_y$ the dispersion
branches are given approximately by
\begin{equation}
E = \hbar v_F  [(\ell\pi/L)^2+k_y^2]^{1/2},
\end{equation}
where $\ell$ is an integer. As a finite magnetic field is
introduced, one can expect a modification of these states. In
particular, one expects the existence of localized states at $k_y
= 0$. This situation is observed in Fig. 4, where the energy
spectrum is plotted as a function of wave vector, for the
potential given by Eq. (6), with $U_0 = 100$ meV, $L = 100$ nm, $W
= 800$ nm and $B = 0.6$ T. This case corresponds to $2\ell_B/L
\approx 0.66$. The figure shows the existence of dispersive states
for a wide range of wavevectors, as well as dispersionless
branches that correspond to the shifted LL at the barriers.

For small energies and wavevectors, the results for $B=0$ show the
existence of hole states in the barriers that either propagate
(Klein paradox) or tunnel through the well. For a finite field, in
a semiclassical description, the trajectories of the holes would
be deflected by the magnetic field and would be confined into
closed orbits that cross the QW. A similar behavior is thus
obtained for energies close to zero, as shown in Fig. 5. The
figure shows the wave functions (left, panels (a) and (c)) and the
respective probability densities (right, panels (b) and (d)) for
two states with $k_y=0$ in a QW with the same parameters as in
Fig. 4 and two energies: $E=2.03$ meV (upper panels) and $E = 0$
(lower panels). The higher energy state describe holes that cross
the well region via electron-hole conversion and thus show a
maximum in the probability density inside the well, whereas the
result for zero energy corresponds to holes that tunnel across the
well region and are confined by the magnetic field. For larger
energy eigenstates and $k_y=0$ the wave functions show an
oscillatory behavior inside the potential well and quickly decay
in the barriers. Therefore, these states possess a dominant
electron-like character.
\begin{figure}
\centering{\resizebox*{!}{9.5cm}{\includegraphics{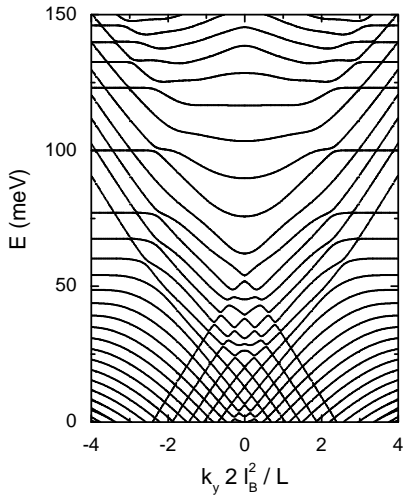}}}
\vspace{-0.5cm} \caption{The same as Fig. 2 but now with $U_0 =
100$ meV, $L = 100$ nm and a smaller magnetic field $B = 0.6$ T.}
\label{fig:f4}
\end{figure}

For finite values of $k_y$ the spectrum displays dispersive
branches with either positive or negative slopes. The wave
functions in these cases have non-negligible amplitudes,
respectively, inside the potential well and in the barriers, and
thus are associated with propagating electrons and holes. On the
other hand, the states associated with flat energy branches have
finite amplitudes inside the barriers. The propagating states are
found to interact with the dispersionless states, with the
appearance of anticrossings.
\begin{figure}
\centering{\resizebox*{!}{8.5cm}{\includegraphics{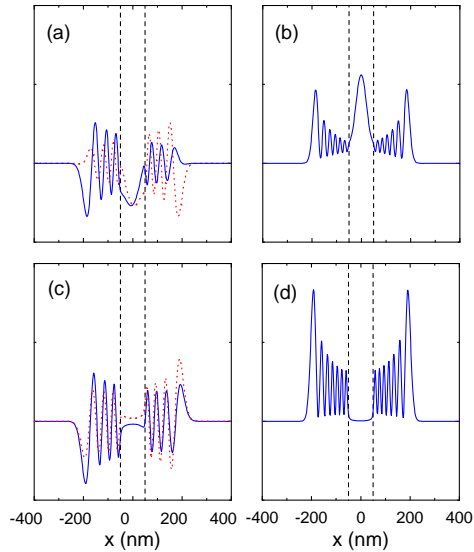}}}
\vspace{-0.5cm} \caption{Color online. Wave functions (left
panels, $\phi_A$: solid, $\phi_B$: dashed-dotted) and the
respective probability densities (right panels) for two low-energy
states of Fig. 4. Upper panels: $E = 2.03$ meV, lower panels:
$E=0$ meV.} \label{fig:f5}
\end{figure}

The right part of Fig. 6 shows comparative plots of $\phi_A$ for
electron sates at the proximity of a specific anticrossing for the
situation corresponding to Fig. 4 and blown up on the left part of
the figure. The results for points (a) and (b) ($k_y 2 \ell_B^2/L=
2.19$) indicate that close to the anticrossings, the electron
states are a superposition of oscillatory states (inside the QW)
and non-propagating states (at the barriers), whereas away from
the anticrossings ($k_y 2 \ell_B^2/L= 2.59$) the wavefunctions
match the results for either confined states (c) in the QW and LL
(d) in the barriers. The vertical lines (dotted) delimit the QW
region.
\begin{figure}
\centering{\resizebox*{!}{8.0cm}{\includegraphics{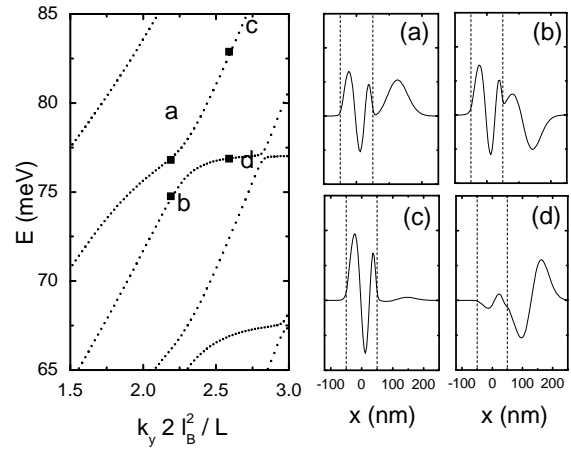}}}
\vspace{-0.5cm} \caption{ {\it Left}: A zoom of the electron
dispersion branches of Fig. 4 at the vicinity of an anticrossing;
the parameters are given in the text. {\it Right} (a-d): Spinor
component $\phi_A$ for some particular states specifed in the
text.}
 \label{fig:f6}
\end{figure}

In the calculation of several physical quantities the density of
states (DOS) is needed, which is defined as
\begin{equation}
\rho(E)= \frac{1}{S}\sum_{k_x,k_y}\sum_n \delta(E-E_n(k_y)),
\end{equation}
where $S = W L$ is the total area of the sample and $n$ labels the
different energy branches. The summation over $k_x$ is replaced by
an integral, and we obtain
\begin{equation}
\rho(E)= \frac{W E_0}{(2\pi \ell_B)^2}\sum_n
\int_{-\infty}^{\infty}\frac{\Gamma/\pi}{(E-E_n(k_y))^2+\Gamma^2}
dk_y,
\end{equation}
where we introduced broadening of the energy levels by replacing
the $\delta$ functions with Lorentzians of constant width
$\Gamma$; $E_0 = \sqrt{2e\hbar B}$ denotes the characteristic
energy scale of the system.

Figure 7 shows numerical results for the DOS for the hyperbolic
tangent QW, with $U_0 = 50$ meV, $L = 200$ nm, and for different
values of the external magnetic field: $B=0.5$ (dashed), $1.0$
(dotted), $1.5$ (dash-dotted) and $2.0$ T (solid). In all cases we
used $\Gamma/E_0 = 28$. The figure shows that the presence of the
barriers shifts the DOS peaks by $U_0$ from the zero potential
results. In contrast with a conventional 2D electron gas, all DOS
results for graphene show a pronounced peak at $E=50$ meV whereas
the remaining peaks are shifted according to the different values
of $B$. This is a consequence of the fact that in graphene, the
energy of the LL with index $n=0$ is independent of $B$. In
addition, for each value of $B$ extra peaks result from the LL
inside the QW. Due to the square root dependence of the spectrum,
see Eq. (5), the distance between the peaks decreases as the
energy increases. Therefore, the influence of the LL in the QW
becomes more evident for energies closer to zero. The inset
contrasts the  DOS of a non-relativistic electron gas (green) with
the DOS of electrons in graphene (solid), for $U_0 = 0$.

The dependence of the DOS on the barrier height $U_0$ is shown in
Fig. 8, where $L = 200$ nm and $B = 2$ T, for different values of
$U_0$, namely $U_0 = 0$ (solid line), $U_0 = 25$ meV (dashed),
$U_0 = 50$ meV (dotted) and $U_0 = 100$ meV (dot-dashed). As the
potential increases, there is a clear shift of the peaks from the
values given by Eq. (5) with $U_0 = 0$, with the $n=0$ peak now
shifted to the value of $U_0$. Also evident is the presence of
states inside the QW, indicated by the existence of additional
peaks, with positions that are independent of the potential step.
These results indicate that, due to the specific nature of the LL
spectrum in graphene, one can increase the DOS at a particular
energy by a suitable change of $U_0$ so that one LL outside the QW
is matched with another LL inside the well. This condition (for
$\Delta = 0$) is expressed as
\begin{equation}
U_0 = \hbar v_F (\sqrt{2n'\beta}-\sqrt{2n\beta}),
\end{equation}
where $n$ and $n'$ are integers. This condition is approached in
Fig. 8 in the result for $U_0 = 25$ meV, for the peak at $E = 84$
meV, with $n' = 4$ and $n = 2$. Figure 9 shows the DOS as a
function of the external magnetic field for $E = 8$ meV, for $U_0
= 0$ (solid line) and $U_0 = 50$ meV (dashed line). The shift of
the LL brought about by the potential barriers causes the
appearance of several DOS peaks that are absent 
from the uniform system.

\begin{figure}
\centering{\resizebox*{!}{9.0cm}{\includegraphics{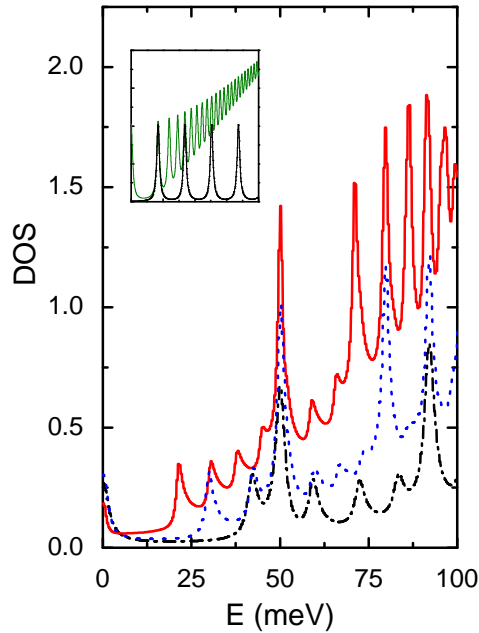}}}
\vspace{-0.5cm} \caption{ Color online. Density of states for a QW
on graphene with $U_0 = 50$ meV, $L = 200$ nm. $B=0.5$ T (dashed),
$B=1.0$ T (dotted), $B=1.5$ T (dash-dotted) and $B=2.0$ T (solid).
Inset: comparison between the DOS of a non-relativistic electron
gas (green) and the DOS of electrons in graphene (solid), for $U_0
= 0$} \label{fig:f7}
\end{figure}

\begin{figure}
\centering{\resizebox*{!}{10.0cm}{\includegraphics{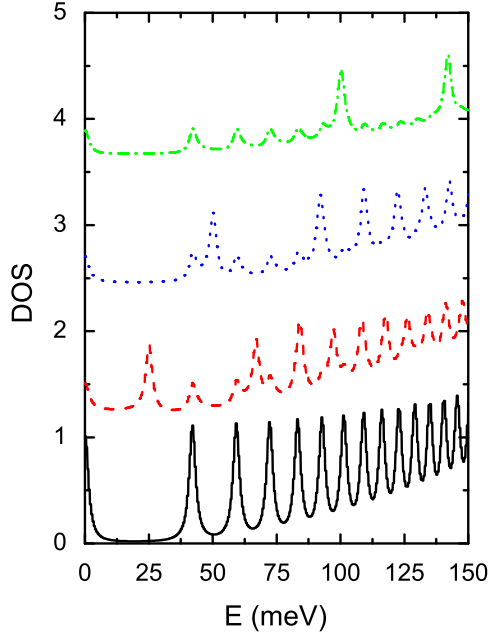}}}
\vspace{-0.5cm} \caption{ Color online. Density of states for a
graphene QW
 with $B = 2$ T and $L = 200$ nm. The solid,
 dashed, dotted,
 and
 dash-dotted curves correspond, respectively, to $U_0=0$,
$U_0=25$ meV , $U_0=50$ meV, and $U_0=100$ meV. The last three
curves are shifted up by $1.25$ for clarity.} \label{fig:f8}
\end{figure}

\begin{figure}
\centering{\resizebox*{!}{8.0cm}{\includegraphics{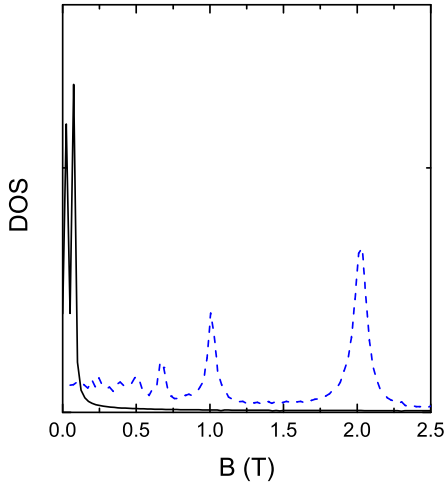}}}
\vspace{-0.5cm} \caption{ Color online. Density of states as a
function of the magnetic field for $E=8$ meV, in the absence of a
confining potential (black solid curve) and 
a graphene QW with $U_0 = 50$ meV (blue dashed curve), $L = 200$
nm in both cases.} \label{fig:f9}
\end{figure}

\subsection{Energy spectrum versus QW parameters}

Figure 10 shows the spectrum, as a function of the magnetic field,
for two values of the wave vector, $k_y = 0$ (a) and $k_y = 0.1$
nm$^{-1}$ (b), for the QW potential of Eq. (6) with $U_0 = 100$
meV and $L = 100$ nm.
\begin{figure}
\centering{\resizebox*{!}{9.0cm}{\includegraphics{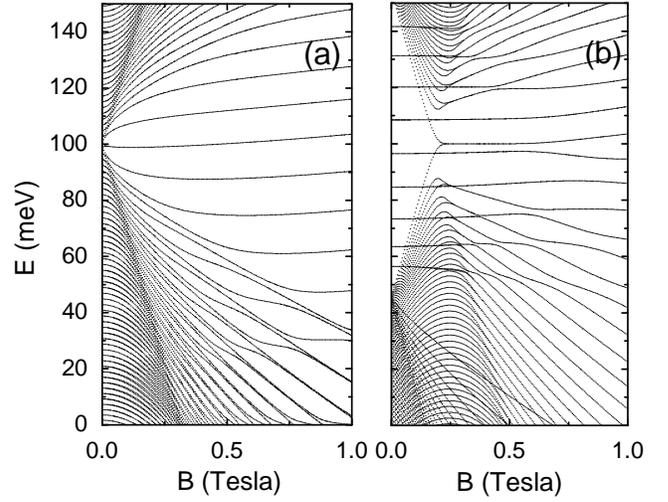}}}
\vspace{-0.5cm} \caption{Energy spectrum of a graphene QW as a
function of the magnetic field $B$, with $U_0 = 100$ meV, $L =
100$ nm, (a) $k_y = 0$, and (b) $k_y = 0.1$ nm$^{-1}$.}
\label{fig:f10}
\end{figure}
\begin{figure}
\centering{\resizebox*{!}{9.0cm}{\includegraphics{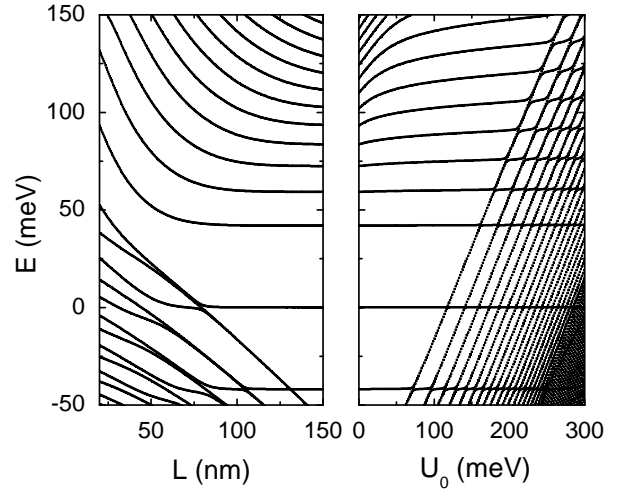}}}
\vspace{-0.5cm} \caption{ Energy spectrum of a graphene QW for
$B=2$ T and $k_y = 0$. In the left panel the spectrum is plotted
vs the width $L$, for $U_0 = 100$ meV, and the right panel vs the
potential height $U_0$ for $L = 100$ nm.}\label{fig:f11}
\end{figure}
In both cases, the results show that the presence of the potential
barriers introduces a significant modification of the energy
eigenstates, in comparison with the results for $U_0=0$ (in which
case the eigenvalues are
 proportional to the square root of the
external magnetic field). Figure 10(b) shows the presence of the
quantized confined states as $B$ tends to 0. These discrete states
are initially very weakly dependent on the external field. As the
field increases, these states interact with the non-propagating
states, as evidenced by the presence of anticrossings. At small
fields the figure shows split branches, which are edge states that
arise due to the fact that the spinor amplitudes are set to zero
at the edges of the sample, whose width is $W = 800$ nm. This
corresponds to the boundary conditions of a graphene sheet with an
armchair edge.

The left panel in Fig. 11 shows the dependence of the energy
eigenvalues on the QW width, for $k_y = 0$, $B = 2$ T and $U_0 =
100$ meV. The results show that the higher-energy levels are
strongly modified even for wide wells, whereas the lower-lying
states can be accurately described by Eq. (5) with $U_0 = 0$ for a
relatively thin QW. The figure allows us to distinguish two
regions in the spectrum, which arise due to the non-uniform
distribution of the LL in graphene. 
One region corresponds to the lower-energy states, which arise due
to the strong interaction between the low-energy states in the
well with the shifted negative-energy states in the barriers. The
degeneracy of these states is lifted for small $L$ and the
spectrum shows an approximately linear dependence on the QW width.
In particular, the interaction with the lower-energy states
modifies the $E=0$ LL for $L \approx 80$ nm, which is equivalent
to $2\ell_B /L \approx 0.45$. The other region corresponds to
higher-energy states, which remain degenerate but are strongly
shifted for small $L$. These states are weakly dependent on the
well width for larger values of $L$, e.g., for $L > 120$ nm.

The dependence of the energy spectrum on the potential barrier
height $U_0$ is shown in the right panel of Fig. 11 for $L = 100$
nm, $B = 2$ T, and $k_y = 0$. As $U_0$ increases, the figure again
shows the appearance of two sets of states: one comprises states
that are weakly dependent on $U_0$ and the other states that show
a significant dependence on $U_0$. This is caused by the hole
states in the barriers, whose energies are shifted by the
potential. As $U_0$  increases, the LL in the well region interact
with the set of shifted of LL in the barriers, causing the
appearance of additional electron states at energies close to
zero.

\section{Summary}
In this work we showed the effect of a confining 1D electrostatic
potential on the energy spectrum of electrons in a graphene QW 
in the presence of a perpendicular magnetic field. We found a
shift of the Landau levels caused by this potential which may be
observable by its effect on the quantum Hall steps in the presence
of gate voltages. For steep potential barriers, there is a clear
distinction between a low magnetic-field regime, characterized by
the existence of dispersive confined states with small wave
vectors inside the QW as well as non-propagating states in the
barriers, and a higher magnetic-field regime, in which there are
non-propagating states inside the well and outside the QW,
together with propagating states at the interfaces of the
potential barriers. These interface states may cross the Fermi
level and, together with the conventional edge states that arise
due to the finite size of the sample, can contribute, e. g., to
the conductivity of the system. The modification of the LL
spectrum in the QW is also evident in the shift of the peaks in
the DOS in comparison with the results for zero confining
potential, along with the appearance of additional peaks caused by
the LL inside the QW.

\section{Acknowledgements}
This work was supported by the Brazilian Council for Research
(CNPq), the Flemish Science Foundation (FWO-Vl), the Belgian
Science Policy (IUAP) and the Canadian NSERC Grant No. OGP0121756.


\begin{thebibliography}{9}

\bibitem{Novo3}
K. S. Novoselov, A. K. Geim, S. V. Morozov, D. Jiang, Y. Zhang, S.
V. Dubonos, I. V. Grigorieva, A. A. Firsov, Science, {\bf 306},
666 (2004).

\bibitem{Novo2}
K. S. Novoselov, D. Jiang, F. Schedin, T. J. Booth, V. V.
Khotkevich, S. V. Morozov, A. K. Geim, PNAS {\bf 102}, 10451
(2005).

\bibitem{Zhang1}
Y. Zhang, J. P. Small, W. V. Pontius, and P. Kim, Appl. Phys.
Lett. {\bf 86}, 073104 (2005).

\bibitem{Zheng}
Y. Zheng and T. Ando, Phys. Rev. B {\bf 65}, 245420 (2002).

\bibitem{Sharapov}
V. P. Gusynin and S. G. Sharapov, Phys. Rev. Lett. {\bf 95},
146801 (2005).

\bibitem{Novoselov}
K. S. Novoselov, A. K. Geim, S. V. Morozov, D. Jiang, M. I.
Katsnelson, I. V. Grigorieva, S. V. Dubonos, A. A. Firsov, Nature
(London) {\bf 438}, 197 (2005).

\bibitem{Zhang}
Y. Zhang, Y. W. Tan, H. L. Stormer, P. Kim, Nature (London) {\bf
438}, 201 (2005).

\bibitem{Abanov}
D. A. Abanin, P. A. Lee and L. S. Levitov, Phys. Rev. Lett. {\bf
96}, 176803 (2006).

\bibitem{Klein}
O. Klein, Z. Phys. {\bf 53}, 157 (1929).

\bibitem{Milton}
J. Milton Pereira Jr., V. Mlinar, F. M. Peeters and P.
Vasilopoulos, Phys. Rev. B {\bf 74}, 045424 (2006).

\bibitem{kat} M. I. Katsnelson, K. S. Novoselov, A. K. Geim, Nature Phys.
{\bf 2}, 620 (2006).

\bibitem{Falko}
V. V. Cheianov and V. I. Fal'ko, Phys. Rev. B {\bf 74}, 041403
(2006).

\bibitem{Milt2}
J. Milton Pereira Jr., V. Mlinar, F. M. Peeters and P.
Vasilopoulos, (unpublished).

\bibitem{Berger}
C. Berger, Z. Song, X. Li, X. Wu, N. Brown, C. Naud, D. Mayou, T.
Li, J. Hass, A. N. Marchenkov, E. H. Conrad, P. N. First and W. A.
de Heer, Science {\bf 312}, 1191 (2006).

\bibitem{Brey}
L. Brey, H. A. Fertig, Phys. Rev. B {\bf 73}, 195408 (2006).

\bibitem{Zhang2}
Y. Zhang, J. P. Small, M. E. S. Amori and P. Kim, Phys. Rev. Lett.
{\bf 94}, 176803 (2005).

\bibitem{High}
J. Reinhardt and W. Greiner, Rep. Prog. Phys. {\bf 40}, 219
(1977); V. Petrillo and Davide Janner, Phys. Rev. A {\bf 67},
012110 (2003).

\bibitem{Wallace}
P. R. Wallace, Phys. Rev. {\bf 71}, 622 (1947); M. Wilson, {\it
Physics Today}, January 2006, p. 21.

\bibitem{Semenoff}
G. W. Semenoff, Phys. Rev. Lett. {\bf 53}, 2449 (1984).

\bibitem{Kopele}
I. A. Luk'yanchuk and Y. Kopelevich, Phys. Rev. Lett. {\bf 93},
166402 (2004).

\bibitem{Kane}
C. L. Kane and E. J. Mele, Phys. Rev. Lett. {\bf 95}, 226801
(2005).


\end{thebibliography}
\end{document}